\documentclass[aps,pra,showpacs,twocolumn,longbibliography]{revtex4-2}

\usepackage{graphicx}
\usepackage{dcolumn}
\usepackage{bm}
\usepackage{color}
\usepackage{amsmath}
\usepackage{amssymb}

\usepackage[colorlinks,citecolor=red,urlcolor=red,bookmarks=false,hypertexnames=true]{hyperref}

\begin{document}

\title{Optimized design of the lithium niobate for spectrally-pure-state generation at MIR wavelengths using metaheuristic algorithm}


\author{Wu-Hao Cai$^{1,a}$}
\author{Ying Tian$^{1,a}$}
\author{Shun Wang$^{1}$}
\author{Chenglong You$^{2}$}
\email{cyou2@lsu.edu}
\author{Qiang Zhou$^{3}$}
\email{zhouqiang@uestc.edu.cn}
\author{Rui-Bo Jin$^{1, 4}$}
\email{jrbqyj@gmail.com}

\date{\today}

\affiliation{$^{1}$ Hubei Key Laboratory of Optical Information and Pattern Recognition, Wuhan Institute of Technology, Wuhan 430205, PR China}
\affiliation{$^{2}$ Quantum Photonics Laboratory, Department of Physics and Astronomy, Louisiana State University, Baton Rouge, LA, 70803 USA}
\affiliation{$^{3}$ Institute of Fundamental and Frontier Sciences and School of Optoelectronic Science and Engineering,\\ University of Electronic Science and Technology of China, Chengdu 610054, China}
\affiliation{$^{4}$ Guangdong Provincial Key Laboratory of Quantum Science and Engineering, Southern University of Science and Technology, Shenzhen 518055, China}
\affiliation{$^{a}$ These authors contributed equally to this work.}


\begin{abstract}
Quantum light sources in the mid-infrared (MIR) band play an important role in many applications, such as quantum sensing, quantum imaging, and quantum communication. However, there is still a lack of high-quality quantum light sources in the MIR band, such as the spectrally pure single-photon source. In this work, we present the generation of spectrally-pure state in an optimized poled lithium niobate crystal using a metaheuristic algorithm. In particular, we adopt the particle swarm optimization algorithm to optimize the duty cycle of the poling period of the lithium niobate crystal. With our approach, the spectral purity can be improved from 0.820 to 0.998 under the third group-velocity-matched condition, and the wavelength-tunable range is from 3.0 $\mu$m to 4.0 $\mu$m for the degenerate case and 3.0 $\mu$m to 3.7 $\mu$m for the nondegenerate case. Our work paves the way for developing quantum photonic technologies at the MIR wavelength band.
\end{abstract}

\maketitle

\section{Introduction}
The single-photon source and entangled photon source at mid-infrared (MIR) wavelength range (approximately 2-20 $\mu$m) are of great interest to many applications, such as quantum sensing, quantum imaging, quantum communication, and quantum measurement \cite{Tournie2019book, Ebrahim-Zadeh2008book,MaganaLoaiza2019,Jin2021MIR}. For example, a single-photon source in the MIR band is useful for gas sensing \cite{Shamy2020SR} and environmental monitoring \cite{Chen2020OLE} with high sensitivity.
Moreover, it can be used for medical imaging \cite{Fernandez2005}, biological sample imaging \cite{Shi2019NP}, and thermal imaging \cite{Godoy2017BOE} with ultra-low light levels. Last but not least, the entangled photon source at MIR wavelength can be used for free-space quantum communication \cite{Bellei2016OE} and for quantum LiDAR with higher precision \cite{Wang2016OE}.

Spontaneous parametric down-conversion (SPDC) in periodically poling lithium niobate (PPLN) is one of the widely used methods to generate MIR-band biphotons, which can be further utilized to produce single-photon source \cite{Sun2019, Jin2020BBO} and entangled photon source. PPLN has the following three advantages for the biphotons generation at MIR wavelengths. Firstly, PPLN has a large nonlinear coefficient and wide transparency range \cite{Kong2020AM}. Secondly, the group-velocity-matched (GVM) wavelengths of LN crystal are intrinsically at the MIR wavelengths \cite{Wei2021}.
Thirdly, PPLN plays an important role in integrated quantum optics \cite{Wang2020PPLN3D}, as it possesses the characteristic of scalability and microminiaturization, which is beneficial in waveguide substrate \cite{Wang2019, Niu2020APL, Lu2022} and thin-film material \cite{Duan2020, Xie2021}.
Up to now, many theoretical and experimental works have utilized PPLN as the medium to prepare the single-photon source \cite{Sua2017SR, Mancinelli2017NC, McCracken2018, Wei2021} or entangled photon source \cite{Prabhakar2020SA} at MIR wavelengths. Recently we theoretically proposed that mid-infrared spectrally-uncorrelated biphotons generation from doped LN crystals, and we find the doping ratio can be utilized as a degree of freedom to manipulate the biphoton state \cite{Wei2021}.

\begin{figure*}[!tbp]
\centering
\includegraphics[width=0.85\textwidth]{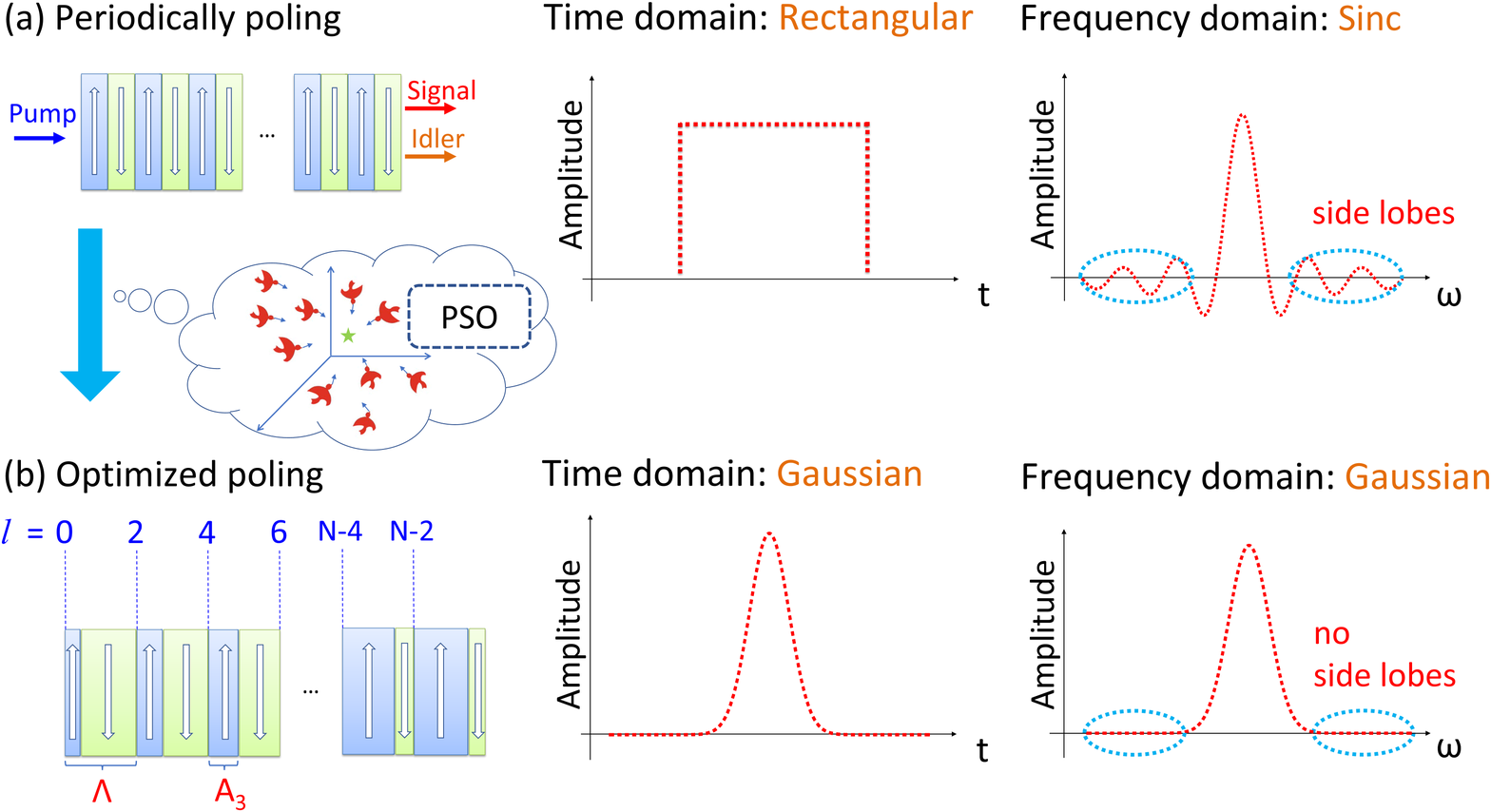}
\caption{The concept of optimizing the poling period of lithium niobate (LN) using particle swarm optimization (PSO) algorithm. (a) In a standard periodically poled LN (PPLN) crystal, the signal and idlers have a rectangular distribution in the time domain and \emph{sinc} distribution in the frequency domain. (b) In an optimized poled LN (OPLN) crystal which is designed using a PSO algorithm, the signal and idler photons have a Gaussian distribution in both time and frequency domains. The side lobes in (a) can be eliminated after optimization, as shown in (b). The number of domains is N, and the number of poling periods is N/2. $\Lambda$ represents the poling period and $A_3$ is the duty cycle at the $3^{rd}$ poling period.}
\label{fig:1}
\end{figure*}

However, PPLN crystals investigated in the previous research are still not optimum due to the low purity of the generated biphotons. The low purity of these biphotons will pose severe restrictions for many quantum applications such as boson sampling, quantum teleportation, and quantum key distribution. \cite{Walmsley2005,Broome2013,Valivarthi2016np,Zhou2017QST,Qi2019,Kwek2021}.
In particular, the current reported best purity under the $3^{rd}$ GVM condition is around 0.82, where the degradation is mainly caused by the phase-matching condition in PPLN \cite{Wei2021}.
As shown in Fig. \ref{fig:1}(a), the single photons generated from the standard PPLN crystal have a rectangular shape in the time domain and \emph{sinc} shape in the frequency domain \cite{Jin2018PRAppl, Jin2021APLphoton}. The side lobes in the \emph{sinc} profile degraded the purity.
In order to improve the purity, the poling period can be designed so that the biphotons have a Gaussian profile in both time and frequency domains. As a result, there are no side lobes in the frequency domain, as illustrated in Fig. \ref{fig:1}(b).

To realize a Gaussian-profiled phase-matching function, several methods have been proposed for the periodically poled potassium titanyl phosphate (PPKTP) crystal at 1550 nm previously. The first one is to modulate the poling order, i.e., lower-order poling in the middle and higher-order poling at the sides of the crystal \cite{Branczyk2011, Kaneda2021}. The second one is to modulate the duty cycle with a near-ideal Gaussian error function \cite{Dixon2013, Chen2019OE}, which could be realized using the machine-learning framework method \cite{Cui2019PRAppl}. Noted that machine learning technology is developing in many fields, such as both classical device \cite{Kim2020} and quantum ones \cite{Zhang2020, Wei2022}. The third one is to modulate the domain sequence using two-domain blocks \cite{Tambasco2016} or one-domain blocks \cite{Graffitti2017, Graffitti2018Optica, Graffitti2020PRL, Graffitti2020PRRes, Pickston2021, Morrison2022}, and this method has also been optimized using a simulated annealing algorithm \cite{Dosseva2016, Quesada2018PRA}. These methods are very effective but limited to the optimization of the KTP crystal.

In our work, we propose a new method for designing optimized poled LN (OPLN) crystal using a metaheuristic algorithm. In particular, we utilize the particle swarm optimization (PSO) algorithm to optimize the duty cycle of the poling period of the LN crystal. The usage of PSO significantly speeds up the optimization process, where thousands of duty cycles are considered. The optimized LN crystal can generate signal and idler photons with Gaussian distributions in both time and frequency domains. Consequently, the spectral purity can be improved to 0.998 under the $3^{rd}$ GVM condition, and the wavelength-tunable range is from 3.0 $\mu$m to 4.0 $\mu$m for the degenerate case and 3.0 $\mu$m to 3.7 $\mu$m for the nondegenerate case. We expect our method will play as a practical tool in crystal structure design and a versatile role in tailoring high-order time-frequency mode. Our OPLN crystal will be great importance for quantum applications in the mid-infrared.

\section{Theory}

\begin{figure}[!t]
\centering\includegraphics[width=0.3\textwidth]{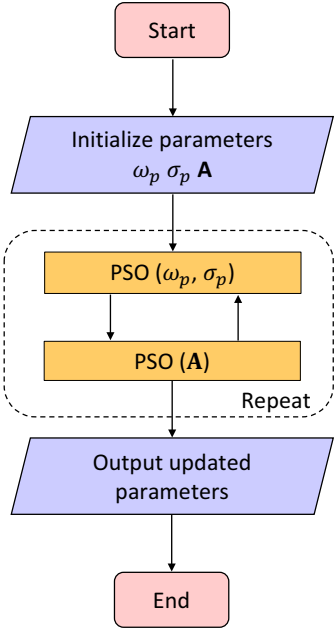}
\caption{The flowchart of our optimization algorithm using PSO. The parameters to be optimized include $\omega_p$: the angular frequency of the pump, $\sigma_p$: the bandwidth of the pump, and $\textbf{A}$: the duty cycles for poling period. After initializing the parameters, the program optimizes the pump parameters $\omega_p$, $\sigma_p$ and the crystal parameters $\textbf{A}$ repeatedly. Finally, the program outputs all the optimized parameters until a stopping criterion is reached.}
\label{fig:2}
\end{figure}

\begin{figure}[!t]
\centering\includegraphics[width=0.48\textwidth]{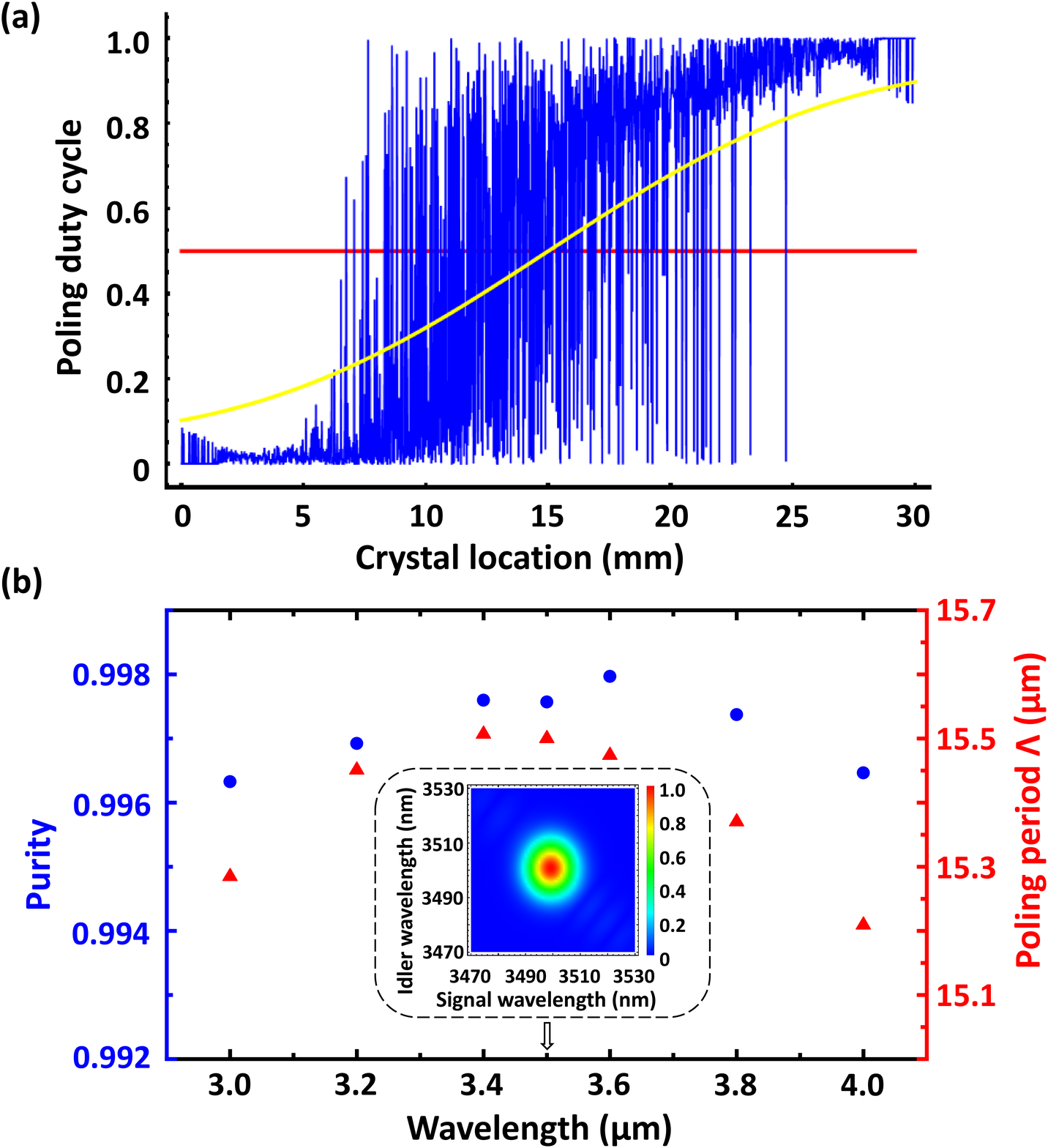}
\caption{(a) The  duty cycle at different crystal positions for 3.5 $\mu$m PSO optimization. The yellow line indicates the  initial parameters of $\textbf{A}$ in the PSO optimization. The blue line is the duty cycle after PSO optimization. (b) The optimized purity and poling period for wavelength-degenerated SPDC for wavelengths from 3.0 $\mu$m to 4.0 $\mu$m.  The inset plot shows the JSA of biphoton state at 3.5 $\mu$m after PSO optimization. } \label{fig:3}
\end{figure}

The conceptual schematic of our approach is depicted in Fig. \ref{fig:1}. We optimize the poling period of the LN crystal to generate signal and idler photons with Gaussian distributions in both time and frequency domains. The biphoton state $\vert\psi\rangle$ generated from an SPDC process can be expressed as

\begin{equation}\label{eq:1}
\vert\psi\rangle=\int_0^\infty\int_0^\infty\,\mathrm{d}\omega_s\,\mathrm{d}\omega_if(\omega_s,\omega_i)\hat a_s^\dag(\omega_s)\hat a_i^\dag(\omega_i)\vert0\rangle_s\vert0\rangle_i,
\end{equation}
where $\omega_{s(i)}$ is the angular frequency of the signal (idler) photon, and $\hat a^\dag$ is the creation operator. In addition, $f(\omega_s,\omega_i)$ is the joint spectral amplitude (JSA) of biphoton, which can be calculated as the product of the pump envelope function (PEF) and phase-matching function (PMF), i.e, $f(\omega_s,\omega_i) = \alpha(\omega_s,\omega_i) \times \phi(\omega_s,\omega_i)$.

The Gaussian PEF $\alpha(\omega_s, \omega_i)$ can be written as
\begin{equation}\label{eq:2}
\alpha(\omega_s, \omega_i)=\exp\left[-\frac{1}{2}\left(\frac{\omega_s+\omega_i-\omega_{p_0}}{\sigma_p}\right)^2\right],
\end{equation}
where $\omega_{p_0} $ and $\sigma_p$ are the central frequency and the bandwidth of the pump. If we choose wavelength as variable by $\omega  = 2\pi c/\lambda$ ($c$ is the light speed in vacuum) for the ease of calculation, the bandwidth of the pump will be described as $\Delta \lambda $. (See Appendix A for more details about PEF and its full width at half-maximum (FWHM)).

Without loss of generality, we assume the crystal is periodically poled along $z$ direction, then the PMF can be expressed as
\begin{equation}\label{eq:3}
\phi(\lambda_s, \lambda_i) = \frac{1}{L}\int_0^L d zg(z)\exp ( - i\Delta kz),
\end{equation}
where $L$ is the length of the crystal and $g(z)=\{+1, -1\}$ indicates the poling configuration orientations. In addition, $\Delta k =k_p-k_i-k_s-2\pi/\Lambda$ is the wave vector mismatch, where $\Lambda$ is the poling period of the crystal, $k_{p(s,i)}=2 \pi n (\lambda_{p(s,i)})/\lambda_{p(s,i)}$ is the wave vector of pump (signal, idler) and $n (\lambda_{p(s,i)})$ is the refractive index as a function of the wavelength $\lambda_{p(s,i)}$.

As illustrated in Fig. \ref{fig:1}(b), the crystal is divided into $N$ domains, and the number of the poling periods is $N/2$. Therefore, $g(z)$ can be rewritten as a function of unit step function $H(z)(H(z)=1, z>0; H(z)=0, z<0)$,

\begin{equation}\label{eq:4}
\begin{aligned}
g(z)=\sum_{l=0}^{N-1}(-1)^{l}\left[H\left(z-z_{l}\right)-H\left(z-z_{l+1}\right)\right],
\end{aligned}
\end{equation}
where $z_{l} = \Lambda l$, $z_{l+1} = \Lambda (l + A_{l}$). In addition, $l\in [0, (N-1)]$ is the domain position, and $A_{l}$ is the duty cycle at the $(l+1)^{th}$ poling period, which constitute the optimized duty cycle matrix $\textbf{A}=\{A_{l}\}$.

By substituting Eq. (\ref{eq:4}) into Eq. (\ref{eq:3}), then PMF can be obtained as
\begin{equation}\label{eq:5}
\begin{aligned}
\phi\left(\lambda_{s}, \lambda_{i}\right)=& \frac{2}{\Delta k L} \sum_{l=0}^{N-1}(-1)^{l}\left\{\sin \left[\Delta k\left(z_{l+1}-z_{l}\right) / 2\right]\right.\\
&\left.e^{\left[-i \Delta k\left(z_{l+1}+z_{l}\right) / 2\right]}\right\}.
\end{aligned}
\end{equation}

As we described above, the side-lobes should be eliminated to ensure the high purity of JSA. Based on Eq. (\ref{eq:2}) and Eq. (\ref{eq:5}), we can optimize PEF and PMF to increase the purity. More specifically, in this paper, our target is to engineer Gaussian-type PMF  and a matched PEF. In our optimization process, we minimize the cost function
\begin{equation}\label{eq:11}
\begin{aligned}
C(\omega_p, \sigma_p, \textbf{A})=\sum_{m=1}^{N_0}{(\phi(\lambda_{s_{m}},\lambda_{i_{m}})-X_{tgt}(m)))}^{2},
\end{aligned}
\end{equation}
where the target PMF $X_{t g t}(m)$ is an Gaussian-shape function given by
\begin{equation}\label{eq:10}
X_{t g t}(m)= \exp \left[-\frac{1}{2}\left(\frac{\lambda_{p_{m}}-\lambda_{p_{0}}}{\sigma_{p}}\right)\right]^{2}.
\end{equation}
Here, $\lambda_{p_{0}}$ is the target center wavelength, and $\sigma_{p}$ is the target bandwidth of PMF matched with PEF. We note that to achieve efficient calculation, we discretized all functions into $N_0$ parts.

\begin{figure*}[!tbp]
\centering\includegraphics[width= 0.85\textwidth]{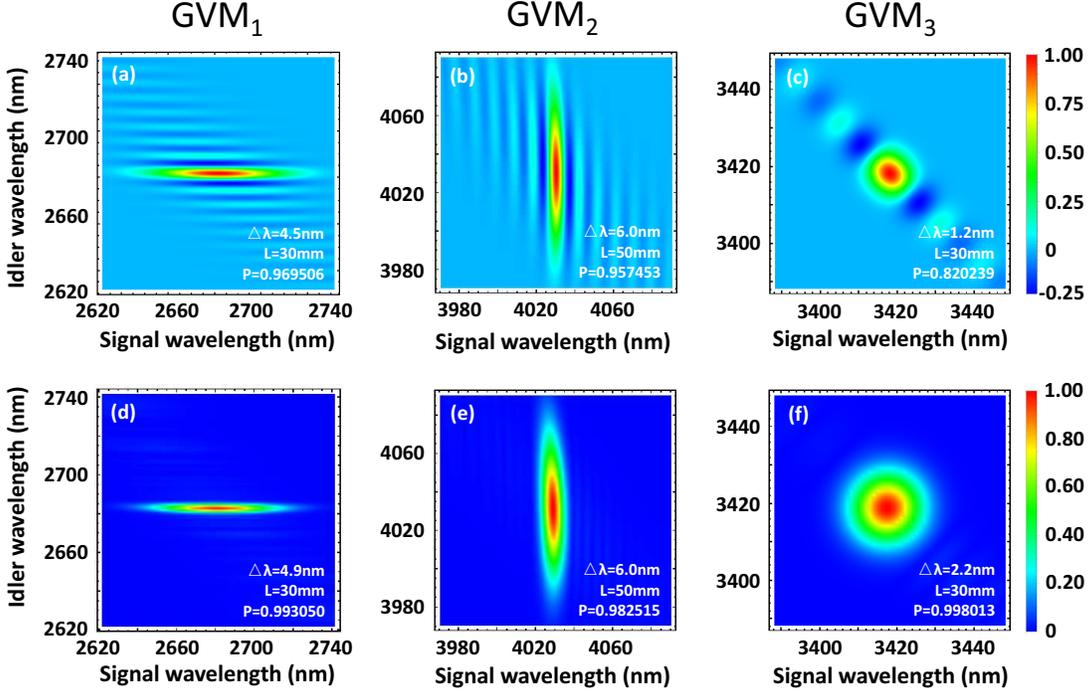}
\caption{
The JSA of the biphoton state is generated from a standard PPLN (a)-(c) and an optimized OPLN (d)-(f). The parameters of pump bandwidth $\Delta\lambda$, the length of the crystal $L$, and the purity $P$ are listed in the figure.  The poling periods of the OPLN are 14.732, 15.181, and 15.504 $\mu$m for $1^{st}(2^{nd},3^{rd})$ GVM cases.
 } \label{fig:4}
\end{figure*}

\section{Optimization\label{sec:3}}

The equations above allow us to utilize the PSO algorithm to optimize the poling structure of the LN crystal. The PSO algorithm is a metaheuristics algorithm that optimizes a function by having a swarm of candidate particles iteratively moving towards the best solutions in the search space \cite{Kennedy1995, Eberhart2001}. We choose the PSO algorithm for this optimization since it can search a very large search space efficiently without using the gradient. This feature is particularly important when the search space scales up, as also in our poling period optimization.

The flowchart of our optimization algorithm is illustrated in Fig. \ref{fig:2}. The overall optimization process is divided into two parts, where two sets of parameters will be optimized: the center frequency $\omega_p$ and bandwidth $\sigma_p$ of the pump laser, and the duty cycle matrix $\textbf{A}$. We first optimize the pump parameters $\omega_p$ and $\sigma_p$ to make sure that $\omega_p$ is at the target center frequency and $\sigma_p$ is matched with the width of the PMF. This step is necessary since $\omega_p$ drifts from the target center frequency during the duty cycle optimization process. We then optimize the duty cycles $\textbf{A}$ of the crystal, to make sure that the PMF has a quasi-Gaussian profile. These two optimization steps are executed iteratively until a stopping criterion is reached.

\section{Results}
\begin{figure*}[!tbp]
\centering\includegraphics[width= 0.9\textwidth]{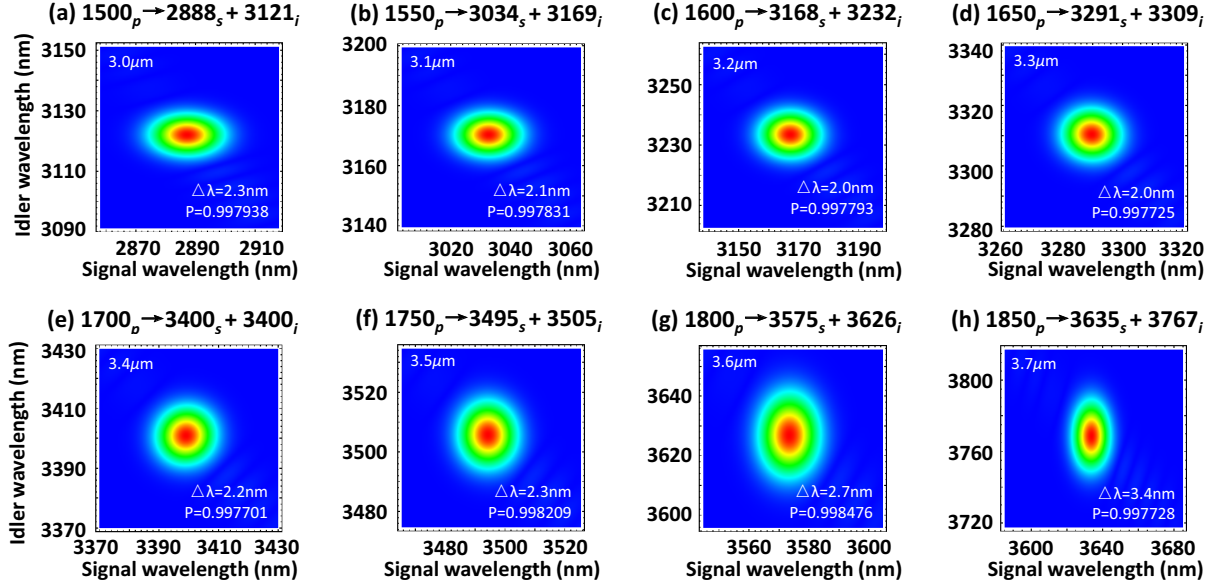}
\caption{
The JSA of the biphoton state for wavelength-nondegenerated SPDC with the duty cycle fixed. The parameters of OPLN at the $3^{rd}$ GVM wavelength 3418nm is from Fig. \ref{fig:4}(f). The pump bandwidth $\Delta\lambda$, and the purity $P$ are listed in the figure.
 } \label{fig:5}
\end{figure*}

First, we show how to optimize the duty cycle of OPLN crystal under wavelength-degenerated condition using our optimization algorithm. We consider a 30-mm-long, type-II (o $\rightarrow$ e+o) phase-matched LN crystal with a wavelength-degenerated case at 3.5 $\mu$m, i.e., 1.75 $\mu$m $\rightarrow$ 3.5 $ \mu$m + 3.5 $\mu$m. We set o-ray as the signal and e-ray as the idler.
In the initialization step, the value of the duty cycle \textbf{A} is set as a Gaussian error function, i.e., the yellow curve in Fig. \ref{fig:3} (a) \cite{Dixon2013}. The center wavelength of the pump is set at 1.75 $\mu$m with bandwidth of 2.3 nm (FWHM = 3.8 nm).
After optimization, we obtain the duty circles of the blue curve shown in Fig. \ref{fig:3} (a). See Appendix B for more optimization details.
%
With the optimized duty cycle and pump parameters, we plot the JSA of biphotons generated from the OPLN, as shown in the inset of Fig. \ref{fig:3} (b).
Then, we can calculate the purity of JSA (see Appendix A for calculation of the purity) and obtain a purity of 0.9975.

Second, following the same procedure, we investigate the tunable range of wavelength-degenerated SPDC at other wavelengths.
We choose the wavelength of 3.0, 3.2, 3.4, 3.5, 3.6, 3.8, and 4.0 $\mu$m to optimize the poling profile.
Optimization results show that the purity can maintain over 0.996 from 3.0 $\mu $m to 4.0 $\mu $m, as indicated by the blue points in Fig. \ref{fig:3}(b), which shows wide tunability at MIR band.
The corresponding poling period is also shown by red points.
For different wavelengths, purity, poling periods, and configurations of pump bandwidth $\Delta\lambda$, FWHM ($\approx$1.67$\Delta\lambda$) and crystal length $L$, are summarized in Table \ref{table:A1} in the Appendix.

Third, we investigate the optimization for other GVM wavelengths, which are important parameters for nonlinear crystals \cite{Edamatsu2011, Jin2019PRAppl}. See Appendix A for the definition of the three GVM conditions.
The three GVM wavelengths calculated for LN crystal are at 2682 nm, 4030 nm, and 3418 nm, respectively. All these GVM wavelengths are located in the MIR band.
Figure \ref{fig:4} (a)-(c) shows the JSA of biphotons generated from a standard PPLN under $1^{st}$ GVM, $2^{nd}$ GVM, and $3^{rd}$ GVM conditions.
The JSAs are distributed along the vertical, horizontal, and diagonal directions, while there are many side lobes are located in the anti-diagonal position. The corresponding purities are 0.97, 0.96, and 0.82, respectively.
After optimization using the algorithm in Fig. \ref{fig:2}, the new JSA for the OPLN can be obtained, as Fig. \ref{fig:4} (d)-(f) depicted.
Comparing Fig. \ref{fig:4}(a)-(c) with (d)-(f), we can find that the side lobes in JSA have been eliminated and the purity of JSA can be improved effectively. In particular, for all cases, the purity has increased to more than 0.98.

Furthermore, we investigate the wavelength non-degenerated case. In practical application, once the OPLN is fabricated, the duty cycle will be fixed. Under this condition, the OPLN can still work with the non-degenerated wavelength. We consider 30-mm-long OPLN designed for degenerated wavelength at 3418 nm (the $3^{rd}$ GVM wavelength), and the poling period is 15.504 $\mu$m.

Figure \ref{fig:5} shows the purity of JSA can be maintained over 0.998
when the pump wavelength at 1.5-1.85 $\mu$m.
The JSA is changing from an oval shape on horizontal position to a perfectly round shape from 3.0 $\mu$m to 3.4 $\mu$m, and then it is changing from a perfectly round shape to an oval shape on the vertical position from 3.4 $\mu$m to 3.7 $\mu$m.

Finally, we note that temperature is also an important parameter for quasi-phase-matching (QPM) crystals since the thermal $-$ optical effect can be used as a tool for precise control of quantum states in many applications \cite{BasiriEsfahani2015}. A short discussion about the performance of thermal properties of optimized OPLN at MIR wavelengths is included in Appendix C.

\section{Discussion and Conclusion}

Recently, machine learning framework Adam has been used to optimize the structure of PPKTP crystal \cite{Cui2019PRAppl}. This study noted that a better resolution needs more computational resources. It means that this method cannot be applied to our work, which has an extremely large parameter space. Thousands of domain and hundreds of wavelength range pieces of JSA involved will render the previous methods unusable. In particular, the machine learning framework approach will be time-consuming, due to its requirement of gradient solving process and huge demand for memory. In turn, we choose the PSO algorithm for this multiparameter optimization, which could speed up without time-consuming gradient solving. We can complete the optimization of PPKTP crystal by PSO in 1 hour while machine learning needs 12 hours or even more. The optimization of PPLN crystal can be accomplished by PSO in around 8$-$12 hours while machine learning algorithm cannot handle this problem (See the detailed configuration of our computer in Appendix). We have depicted cost variation in the Appendix section, the speed of convergence is very fast. Therefore, the convergence shows the feasibility of our work, meanwhile, it shows that PSO is a versatile tool in the crystal structure design.
In addition, we believe our method can be utilized for designing special non-linear crystals to generate exotic quantum states for quantum sensing and metrology \cite{Graffitti2020PRL, Jin2021APLphoton, Morrison2022, Jin2016QST}.
We note that other optimization algorithms, such as simulated annealing or genetic algorithms, are also suitable for this kind of optimization problem.





Note in the calculation of purity in the main text (e.g., in Fig \ref{fig:3}, Fig \ref{fig:4}, Fig \ref{fig:5}), the calculation range is 60 nm. This is equivalent to applying a 60-nm-wide square-shaped bandpass filter on the source.
We also considered the calculation at a wider range up to 240 nm in Appendix D. The calculated purities are 0.998, 0.996, and 0.973 when the wavelength range are 60 nm, 120 nm, and 240 nm. This means the purity does not decrease significantly using BPF with a larger width.

In summary, we have proposed an optimization scheme of PPLN crystal for the spectrally pure-state generation. By using the metaheuristic algorithm, we can generate the pure-state with an extremely high purity of 0.998 at the $3^{rd}$ GVM wavelength. In addition, we demonstrate our method could be extended to the wide tunable range and maintain the purity above 0.996 from 3.0 $\mu$m to 4.0 $\mu$m in the degenerate case. Using the designed OPLN crystal, the purity will stay at around 0.998 from 3.0 $\mu$m to 3.7 $\mu$m in the nondegenerate case. The optimization under the $1^{st}$ and $2^{nd}$ GVM conditions and the thermal properties are also discussed. Our work is important for quantum technology development in the MIR band, and our optimization method may be a versatile tool in crystal structure design in MIR applications.



\section*{Acknowledgments}
We thank Prof. Zheshen Zhang and Dr. Chaohan Cui for the helpful discussion. This work is supported by the National Natural Science Foundations of China (Grant Nos.12074299, 91836102, 11704290) and by the Guangdong Provincial Key Laboratory (Grant No. GKLQSE202102).

\section*{Conflict of Interest}
The authors declare that there are no conflicts of interest related to this article.

\bibliographystyle{osajnl}

\begin{thebibliography}{10}
\newcommand{\enquote}[1]{``#1''}

\bibitem{Tournie2019book}
E.~Tournie and L.~Cerutti, eds., \emph{Mid-infrared optoelectronics materials,
  devices, and applications} (Woodhead Publishing, 2019).

\bibitem{Ebrahim-Zadeh2008book}
M.~Ebrahim-Zadeh and I.~T. Sorokina, eds., \emph{Mid-Infrared Coherent Sources
  and Applications} (Springer Science \& Business Media, 2008).

\bibitem{MaganaLoaiza2019}
O.~S. Maga{\~{n}}a-Loaiza and R.~W. Boyd, \enquote{Quantum imaging and
  information,} Rep. Prog. Phys. \textbf{82}, 124401 (2019).

\bibitem{Jin2021MIR}
R.-B. Jin and Y.~Tian, \enquote{{Research progress on the mid-infrared band
  quantum light source (in Chinese)},} {Journal of Anhui University (Natural
  Science Edition)} \textbf{45}, 10--19 (2021).

\bibitem{Shamy2020SR}
R.~S.~E. Shamy, D.~Khalil, and M.~A. Swillam, \enquote{Mid infrared optical gas
  sensor using plasmonic {Mach-Zehnder} interferometer,} Sci. Rep. \textbf{10},
  1293 (2020).

\bibitem{Chen2020OLE}
K.~Chen, S.~Liu, B.~Zhang, Z.~Gong, Y.~Chen, M.~Zhang, H.~Deng, M.~Guo, F.~Ma,
  F.~Zhu, and Q.~Yu, \enquote{Highly sensitive photoacoustic multi-gas analyzer
  combined with mid-infrared broadband source and near-infrared laser,} Opt.
  Lasers Eng. \textbf{124}, 105844 (2020).

\bibitem{Fernandez2005}
D.~C. Fernandez, R.~Bhargava, S.~M. Hewitt, and I.~W. Levin, \enquote{Infrared
  spectroscopic imaging for histopathologic recognition,} Nat. Biotech.
  \textbf{23}, 469--474 (2005).

\bibitem{Shi2019NP}
J.~Shi, T.~T.~W. Wong, Y.~He, L.~Li, R.~Zhang, C.~S. Yung, J.~Hwang, K.~Maslov,
  and L.~V. Wang, \enquote{High-resolution, high-contrast mid-infrared imaging
  of fresh biological samples with ultraviolet-localized photoacoustic
  microscopy,} Nat. Photonics \textbf{13}, 609--615 (2019).

\bibitem{Godoy2017BOE}
S.~E. Godoy, M.~M. Hayat, D.~A. Ramirez, S.~A. Myers, R.~S. Padilla, and
  S.~Krishna, \enquote{Detection theory for accurate and non-invasive skin
  cancer diagnosis using dynamic thermal imaging,} Biomed. Opt. Express
  \textbf{8}, 2301--2323 (2017).

\bibitem{Bellei2016OE}
F.~Bellei, A.~P. Cartwright, A.~N. McCaughan, A.~E. Dane, F.~Najafi, Q.~Zhao,
  and K.~K. Berggren, \enquote{Free-space-coupled superconducting nanowire
  single-photon detectors for infrared optical communications,} Opt. Express
  \textbf{24}, 3248--3257 (2016).

\bibitem{Wang2016OE}
Q.~Wang, L.~Hao, Y.~Zhang, L.~Xu, C.~Yang, X.~Yang, and Y.~Zhao,
  \enquote{Super-resolving quantum lidar: entangled coherent-state sources with
  binary-outcome photon counting measurement suffice to beat the shot-noise
  limit,} Opt. Express \textbf{24}, 5045--5056 (2016).

\bibitem{Sun2019}
X.~Sun, P.~Wang, B.~Sheng, T.~Wang, Z.~Chen, K.~Gao, M.~Li, J.~Zhang, W.~Ge,
  Y.~Arakawa, B.~Shen, M.~Holmes, and X.~Wang, \enquote{Single-photon emission
  from a further confined {InGaN}/{GaN} quantum disc via reverse-reaction
  growth,} Quantum Eng. \textbf{1}, e20 (2019).

\bibitem{Jin2020BBO}
R.-B. Jin, W.-H. Cai, C.~Ding, F.~Mei, G.-W. Deng, R.~Shimizu, and Q.~Zhou,
  \enquote{{Spectrally uncorrelated biphotons generated from `the family of BBO
  crystal'},} Quantum Eng. \textbf{2}, e38 (2020).

\bibitem{Kong2020AM}
Y.~Kong, F.~Bo, W.~Wang, D.~Zheng, H.~Liu, G.~Zhang, R.~Rupp, and J.~Xu,
  \enquote{Recent progress in lithium niobate: Optical damage, defect
  simulation, and on-chip devices,} Adv. Mater. \textbf{32}, 1806452 (2020).

\bibitem{Wei2021}
B.~Wei, W.-H. Cai, C.~Ding, G.-W. Deng, R.~Shimizu, Q.~Zhou, and R.-B. Jin,
  \enquote{Mid-infrared spectrally-uncorrelated biphotons generation from doped
  {PPLN}: a theoretical investigation,} Opt. Express \textbf{29}, 256--271
  (2021).

\bibitem{Wang2020PPLN3D}
T.~Wang, P.~Chen, C.~Xu, Y.~Zhang, D.~Wei, X.~Hu, G.~Zhao, M.~Xiao, and S.~Zhu,
  \enquote{{Periodically poled LiNbO$_3$ crystals from 1D and 2D to 3D},} Sci.
  China Technol. Sci. \textbf{63}, 1110--1126 (2020).

\bibitem{Wang2019}
M.~Wang, R.~Wu, J.~Lin, J.~Zhang, Z.~Fang, Z.~Chai, and Y.~Cheng,
  \enquote{Chemo-mechanical polish lithography: A pathway to low loss
  large-scale photonic integration on lithium niobate on insulator,} Quantum
  Eng. \textbf{1}, e9 (2019).

\bibitem{Niu2020APL}
Y.~Niu, C.~Lin, X.~Liu, Y.~Chen, X.~Hu, Y.~Zhang, X.~Cai, Y.-X. Gong, Z.~Xie,
  and S.~Zhu, \enquote{Optimizing the efficiency of a periodically poled {LNOI}
  waveguide using in situ monitoring of the ferroelectric domains,} Appl. Phys.
  Lett. \textbf{116}, 101104 (2020).

\bibitem{Lu2022}
C.~Lu, H.~Li, J.~Qiu, Y.~Zhang, S.~Liu, Y.~Zheng, and X.~Chen, \enquote{Second
  and cascaded harmonic generation of pulsed laser in a lithium niobate on
  insulator ridge waveguide,} Opt. Express \textbf{30}, 1381--1387 (2022).

\bibitem{Duan2020}
J.-C. Duan, J.-N. Zhang, Y.-J. Zhu, C.-W. Sun, Y.-C. Liu, P.~Xu, Z.~Xie, Y.-X.
  Gong, and S.-N. Zhu, \enquote{Generation of narrowband counterpropagating
  polarization-entangled photon pairs based on thin-film lithium niobate on
  insulator,} J. Opt. Soc. Am. B \textbf{37}, 2139--2145 (2020).

\bibitem{Xie2021}
R.-R. Xie, G.-Q. Li, F.~Chen, and G.-L. Long, \enquote{Microresonators in
  lithium niobate thin films,} Adv. Opt. Mater. \textbf{9}, 2100539 (2021).

\bibitem{Sua2017SR}
Y.~M. Sua, H.~Fan, A.~Shahverdi, J.-Y. Chen, and Y.-P. Huang, \enquote{Direct
  generation and detection of quantum correlated photons with 3.2 um wavelength
  spacing,} Sci. Rep. \textbf{7}, 17494 (2017).

\bibitem{Mancinelli2017NC}
M.~Mancinelli, A.~Trenti, S.~Piccione, G.~Fontana, J.~S. Dam,
  P.~Tidemand-Lichtenberg, C.~Pedersen, and L.~Pavesi, \enquote{Mid-infrared
  coincidence measurements on twin photons at room temperature,} Nat. Commun.
  \textbf{8}, 15184 (2017).

\bibitem{McCracken2018}
R.~A. McCracken, F.~Graffitti, and A.~Fedrizzi, \enquote{Numerical
  investigation of mid-infrared single-photon generation,} J. Opt. Soc. Am. B
  \textbf{35}, C38--C48 (2018).

\bibitem{Prabhakar2020SA}
S.~Prabhakar, T.~Shields, A.~C. Dada, M.~Ebrahim, G.~G. Taylor, D.~Morozov,
  K.~Erotokritou, S.~Miki, M.~Yabuno, H.~Terai, C.~Gawith, M.~Kues, L.~Caspani,
  R.~H. Hadfield, and M.~Clerici, \enquote{Two-photon quantum interference and
  entanglement at 2.1 $\mu$m,} Sci. Adv. \textbf{6}, eaay5195 (2020).

\bibitem{Walmsley2005}
I.~A. Walmsley and M.~G. Raymer, \enquote{Toward quantum-information processing
  with photons,} Science \textbf{307}, 1733 (2005).

\bibitem{Broome2013}
M.~A. Broome, A.~Fedrizzi, S.~Rahimi-Keshari, J.~Dove, S.~Aaronson, T.~C.
  Ralph, and A.~G. White, \enquote{Photonic boson sampling in a tunable
  circuit,} Science \textbf{339}, 794 (2013).

\bibitem{Valivarthi2016np}
R.~Valivarthi, M.~G. Puigibert, Q.~Zhou, G.~H. Aguilar, V.~B. Verma,
  F.~Marsili, M.~D. Shaw, S.~W. Nam, D.~Oblak, and W.~Tittel, \enquote{Quantum
  teleportation across a metropolitan fibre network,} Nat. Photonics
  \textbf{10}, 676 (2016).

\bibitem{Zhou2017QST}
R.~Valivarthi, Q.~Zhou, C.~John, F.~Marsili, V.~B. Verma, M.~D. Shaw, S.~W.
  Nam, D.~Oblak, and W.~Tittel, \enquote{A cost-effective
  measurement-device-independent quantum key distribution system for quantum
  networks,} Quantum Sci. Technol. \textbf{2}, 04LT01 (2017).

\bibitem{Qi2019}
R.~Qi, Z.~Sun, Z.~Lin, P.~Niu, W.~Hao, L.~Song, Q.~Huang, J.~Gao, L.~Yin, and
  G.-L. Long, \enquote{Implementation and security analysis of practical
  quantum secure direct communication,} Light Sci. Appl \textbf{8}, 22 (2019).

\bibitem{Kwek2021}
L.-C. Kwek, L.~Cao, W.~Luo, Y.~Wang, S.~Sun, X.~Wang, and A.~Q. Liu,
  \enquote{Chip-based quantum key distribution,} {AAPPS} Bull. \textbf{31}, 15
  (2021).

\bibitem{Jin2018PRAppl}
R.-B. Jin, T.~Saito, and R.~Shimizu, \enquote{Time-frequency duality of
  biphotons for quantum optical synthesis,} Phys. Rev. Applied \textbf{10},
  034011 (2018).

\bibitem{Jin2021APLphoton}
R.-B. Jin, K.~Tazawa, N.~Asamura, M.~Yabuno, S.~Miki, F.~China, H.~Terai,
  K.~Minoshima, and R.~Shimizu, \enquote{{Quantum optical synthesis in 2D
  time–frequency space},} APL Photon. \textbf{6}, 086104 (2021).

\bibitem{Branczyk2011}
A.~M. Bra\'nczyk, A.~Fedrizzi, T.~M. Stace, T.~C. Ralph, and A.~G. White,
  \enquote{Engineered optical nonlinearity for quantum light sources,} Opt.
  Express \textbf{19}, 55--65 (2011).

\bibitem{Kaneda2021}
F.~Kaneda, J.~Oikawa, M.~Yabuno, F.~China, S.~Miki, H.~Terai, Y.~Mitsumori, and
  K.~Edamatsu, \enquote{Generation of spectrally factorable photon pairs via
  multi-order quasi-phase-matched spontaneous parametric downconversion,}
  arXiv:2111.10981  (2021).

\bibitem{Dixon2013}
P.~B. Dixon, J.~H. Shapiro, and F.~N.~C. Wong, \enquote{Spectral engineering by
  {Gaussian} phase-matching for quantum photonics,} Opt. Express \textbf{21},
  5879--5890 (2013).

\bibitem{Chen2019OE}
C.~Chen, J.~E. Heyes, K.-H. Hong, M.~Y. Niu, A.~E. Lita, T.~Gerrits, S.~W. Nam,
  J.~H. Shapiro, and F.~N.~C. Wong, \enquote{Indistinguishable single-mode
  photons from spectrally engineered biphotons,} Opt. Express \textbf{27},
  11626--11634 (2019).

\bibitem{Cui2019PRAppl}
C.~Cui, R.~Arian, S.~Guha, N.~Peyghambarian, Q.~Zhuang, and Z.~Zhang,
  \enquote{Wave-function engineering for spectrally uncorrelated biphotons in
  the telecommunication band based on a machine-learning framework,} Phys. Rev.
  Appl. \textbf{12}, 034059 (2019).

\bibitem{Kim2020}
S.~Kim, \enquote{Neuromorphic computing for machine learning acceleration based
  on spiking neural network,} {AAPPS} Bull. \textbf{30}, 21--25 (2020).

\bibitem{Zhang2020}
Y.~Zhang and Q.~Ni, \enquote{Recent advances in quantum machine learning,}
  Quantum Eng. \textbf{2}, e34 (2020).

\bibitem{Wei2022}
S.~Wei, Y.~Chen, Z.~Zhou, and G.~Long, \enquote{A quantum convolutional neural
  network on {NISQ} devices,} {AAPPS} Bull. \textbf{32}, 2 (2022).

\bibitem{Tambasco2016}
J.-L. Tambasco, A.~Boes, L.~G. Helt, M.~J. Steel, and A.~Mitchell,
  \enquote{Domain engineering algorithm for practical and effective photon
  sources,} Opt. Express \textbf{24}, 19616--19626 (2016).

\bibitem{Graffitti2017}
F.~Graffitti, D.~Kundys, D.~T. Reid, A.~M. Bra{\'{n}}czyk, and A.~Fedrizzi,
  \enquote{Pure down-conversion photons through sub-coherence-length domain
  engineering,} Quantum Sci. Technol. \textbf{2}, 035001 (2017).

\bibitem{Graffitti2018Optica}
F.~Graffitti, P.~Barrow, M.~Proietti, D.~Kundys, and A.~Fedrizzi,
  \enquote{Independent high-purity photons created in domain-engineered
  crystals,} Optica \textbf{5}, 514--517 (2018).

\bibitem{Graffitti2020PRL}
F.~Graffitti, P.~Barrow, A.~Pickston, A.~M. Bra\ifmmode~\acute{n}\else
  \'{n}\fi{}czyk, and A.~Fedrizzi, \enquote{Direct generation of tailored
  pulse-mode entanglement,} Phys. Rev. Lett. \textbf{124}, 053603 (2020).

\bibitem{Graffitti2020PRRes}
F.~Graffitti, V.~D'Ambrosio, M.~Proietti, J.~Ho, B.~Piccirillo, C.~de~Lisio,
  L.~Marrucci, and A.~Fedrizzi, \enquote{Hyperentanglement in structured
  quantum light,} Phys. Rev. Research \textbf{2}, 043350 (2020).

\bibitem{Pickston2021}
A.~Pickston, F.~Graffitti, P.~Barrow, C.~L. Morrison, J.~Ho, A.~M.
  Bra\'{n}czyk, and A.~Fedrizzi, \enquote{Optimised domain-engineered crystals
  for pure telecom photon sources,} Opt. Express \textbf{29}, 6991--7002
  (2021).

\bibitem{Morrison2022}
C.~L. Morrison, F.~Graffitti, P.~Barrow, A.~Pickston, J.~Ho, and A.~Fedrizzi,
  \enquote{Frequency-bin entanglement from domain-engineered down-conversion,}
  arXiv: 2201.07259  (2022).

\bibitem{Dosseva2016}
A.~Dosseva, L.~Cincio, and A.~M. Bra\'nczyk, \enquote{Shaping the joint
  spectrum of down-converted photons through optimized custom poling,} Phys.
  Rev. A \textbf{93}, 013801 (2016).

\bibitem{Quesada2018PRA}
N.~Quesada and A.~M. Bra\ifmmode~\acute{n}\else \'{n}\fi{}czyk,
  \enquote{Gaussian functions are optimal for waveguided
  nonlinear-quantum-optical processes,} Phys. Rev. A \textbf{98}, 043813
  (2018).

\bibitem{Kennedy1995}
J.~Kennedy and R.~Eberhart, \enquote{Particle swarm optimization,} {Proceedings
  of ICNN'95 - International Conference on Neural Networks} \textbf{4},
  1942--1948 (1995).

\bibitem{Eberhart2001}
Eberhart and Y.~Shi, \enquote{Particle swarm optimization: developments,
  applications and resources,} Proceedings of the 2001 Congress on Evolutionary
  Computation (IEEE Cat. No.01TH8546) \textbf{1}, 81--86 (2001).

\bibitem{Edamatsu2011}
K.~Edamatsu, R.~Shimizu, W.~Ueno, R.-B. Jin, F.~Kaneda, M.~Yabuno, H.~Suzuki,
  S.~Nagano, A.~Syouji, and K.~Suizu, \enquote{Photon pair sources with
  controlled frequency correlation,} Prog. Inform. \textbf{8}, 19--26 (2011).

\bibitem{Jin2019PRAppl}
R.-B. Jin, N.~Cai, Y.~Huang, X.-Y. Hao, S.~Wang, F.~Li, H.-Z. Song, Q.~Zhou,
  and R.~Shimizu, \enquote{Theoretical investigation of a spectrally pure-state
  generation from isomorphs of {KDP} crystal at near-infrared and telecom
  wavelengths,} Phys. Rev. Appl. \textbf{11}, 034067 (2019).

\bibitem{BasiriEsfahani2015}
S.~Basiri-Esfahani, C.~R. Myers, A.~Armin, J.~Combes, and G.~J. Milburn,
  \enquote{Integrated quantum photonic sensor based on {Hong-Ou-Mandel}
  interference,} Opt. Express \textbf{23}, 16008 (2015).

\bibitem{Jin2016QST}
R.-B. Jin, R.~Shimizu, M.~Fujiwara, M.~Takeoka, R.~Wakabayashi, T.~Yamashita,
  S.~Miki, H.~Terai, T.~Gerrits, and M.~Sasaki, \enquote{Simple method of
  generating and distributing frequency-entangled qudits,} Quantum Sci.
  Technol. \textbf{1}, 015004 (2016).

\bibitem{Jin2013OE}
R.-B. Jin, R.~Shimizu, K.~Wakui, H.~Benichi, and M.~Sasaki, \enquote{Widely
  tunable single photon source with high purity at telecom wavelength,} Opt.
  Express \textbf{21}, 10659--10666 (2013).

\bibitem{Mosley2008NJP}
P.~J. Mosley, J.~S. Lundeen, B.~J. Smith, and I.~A. Walmsley,
  \enquote{Conditional preparation of single photons using parametric
  downconversion: a recipe for purity,} New J. Phys. \textbf{10}, 093011
  (2008).

\bibitem{Jundt1990}
D.~Jundt, M.~Fejer, and R.~Byer, \enquote{Optical properties of lithium-rich
  lithium niobate fabricated by vapor transport equilibration,} IEEE J. Quantum
  Electron. \textbf{26}, 135--138 (1990).

\end{thebibliography}

\setcounter{figure}{0}
\renewcommand\thefigure{A\arabic{figure}}

\setcounter{equation}{0}
\renewcommand\theequation{A\arabic{equation}}

\renewcommand\thetable{A\arabic{table}}

\clearpage
\newpage
\onecolumngrid

\section*{Appendix}

\subsection*{A. Definition of the FWHM, the GVM conditions, and the purity}
Firstly, we show the definition of the FWHM. According to Eq. (\ref{eq:2}), the half-maximum intensity of PEF can be expressed as
\begin{equation}\label{FWHM1}
{\left( {\exp \left[ { - \frac{1}{2}{{\left( {\frac{{{\omega _p} - {\omega _{{p_0}}}}}{{{\sigma _p}}}} \right)}^2}} \right]} \right)^2} = \frac{1}{2},
\end{equation}

The corresponding angular frequency $\omega _{p1}$ and $\omega _{p2}$ at the half-maximum intensity will be
\begin{equation}\label{FWHM2}
\begin{aligned}
{\omega _{p1}} = {\omega _{{p_0}}} - \frac{{\textrm{FWHM}_\omega}}{2},\\
{\omega _{p2}} = {\omega _{{p_0}}} + \frac{{\textrm{FWHM}_\omega}}{2}.
\end{aligned}
\end{equation}

The solution to Eq. (\ref{FWHM1}) are
\begin{equation}\label{FWHM3}
\begin{aligned}
{\omega _{p1}} = {\omega _{{p_0}}} - {\sigma _p}\sqrt{\ln (2)},\\
{\omega _{p2}} = {\omega _{{p_0}}} + {\sigma _p}\sqrt{\ln (2)}.
\end{aligned}
\end{equation}

From Eq. (\ref{FWHM3}), we can obtain the FWHM of the pump as $\text{FWHM}_\omega = 2\sqrt{\ln(2)} \sigma_p  \approx 1.67\sigma_p $.

The PEF also can be rewritten by using wavelength as variable by considering $\omega  = 2\pi c/\lambda$ ($c$ is lightspeed in vacuum),

\begin{equation}\label{PEF}
\begin{aligned}
\alpha\left(\lambda_{s}, \lambda_{i}\right)=\exp \left(-\frac{1}{2}\left\{\frac{1 / \lambda_{s}+1 / \lambda_{i}-1 /\left(\lambda_{0} / 2\right)}{\Delta \lambda /\left[\left(\lambda_{0} / 2\right)^{2}-(\Delta \lambda / 2)^{2}\right]}\right\}^{2}\right),
\end{aligned}
\end{equation}

The bandwidth is given by $\sigma_p = {2\pi c\frac{{\Delta \lambda }}{{{{\left( {{\lambda _0}/2} \right)}^2} - {{\left( {\Delta \lambda /2} \right)}^2}}}}$, where $\lambda_0/2$ is the central wavelength of the pump, $\Delta\lambda$ is the bandwidth of the pump. The full width at half maximum (FWHM) of the pump at intensity level is

\begin{equation}\label{FWHMl}
\begin{aligned}
\text{FWHM}_\lambda= \frac{2\sqrt{\ln (2)} {\lambda _0}^2   \Delta \lambda    \left({\lambda _0}^2-\Delta \lambda ^2\right)}{{\lambda _0}^4+\Delta \lambda ^4-2 {\lambda _0}^2 \Delta \lambda ^2 [1+\ln (4)]}.
\end{aligned}
\end{equation}
For $\Delta \lambda \ll \lambda_0$, $\text{FWHM}_\lambda\approx 2\sqrt{\ln(2)} \Delta \lambda  \approx 1.67\Delta \lambda $.

Next, we explain the definition of the GVM conditions.
The tilt angle $\theta$ of PMF is determined by \cite{Jin2013OE}:
\begin{equation}
\begin{aligned}
\tan \theta=-\left(\frac{V_{g, p}^{-1}\left(\omega_{p}\right)-V_{g, s}^{-1}\left(\omega_{s}\right)}{V_{g, p}^{-1}\left(\omega_{p}\right)-V_{g, i}^{-1}\left(\omega_{i}\right)}\right),
\end{aligned}
\end{equation}
where $V_{g, \mu}=\frac{d \omega}{d k_{\mu}(\omega)}=\frac{1}{k_{\mu}^{\prime}(\omega)},(\mu=p, s, i)$ is the group velocity.

With tilting angle $\theta$ changing, the shape of the PMF is determined by following three GVM conditions:

The $1^{st}$ GVM (GVM$_1$) condition ($\theta=0^\circ$)
\begin{equation}\label{GVM1}
\begin{aligned}
V_{g, p}^{-1}\left(\omega_{p}\right)=V_{g, s}^{-1}\left(\omega_{s}\right),
\end{aligned}
\end{equation}

The $2^{nd}$ GVM (GVM$_2$) condition ($\theta=90^\circ$)
\begin{equation}\label{GVM2}
\begin{aligned}
V_{g, p}^{-1}\left(\omega_{p}\right)=V_{g, i}^{-1}\left(\omega_{i}\right),
\end{aligned}
\end{equation}

The $3^{rd}$ GVM (GVM$_3$) condition ($\theta=45^\circ$)
\begin{equation}\label{GVM3}
\begin{aligned}
2V_{g, p}^{-1}\left(\omega_{p}\right)=V_{g, s}^{-1}\left(\omega_{s}\right)+V_{g, i}^{-1}\left(\omega_{i}\right).
\end{aligned}
\end{equation}

Finally, we show the definition of the purity.

The purity of JSA could be calculated through Schmidt decomposition on $f(\omega_s,\omega_i)$ \cite{Mosley2008NJP}:
\begin{equation}\label{schmidt}
f(\omega_s,\omega_i) = \sum_jc_j\phi_j(\omega_s)\varphi(\omega_i),
\end{equation}
where $\phi_j(\omega_s)$ and $\varphi_j(\omega_i)$ are the two orthogonal basis vectors in the frequency domain, and $c_j$ is a set of non-negative real numbers that satisfy the normalization condition $ \sum_jc_j^2=1$.
The purity $P$ is defined as:
\begin{equation}\label{purity}
P=\sum_jc_j^4.
\end{equation}

Note that for all the calculation of purity in this study, we use a grid size of 200 $\times$ 200 for all the JSAs.
\subsection*{B. Optimization details}
\begin{figure*}[htb]
\centering\includegraphics[width= 0.5\textwidth]{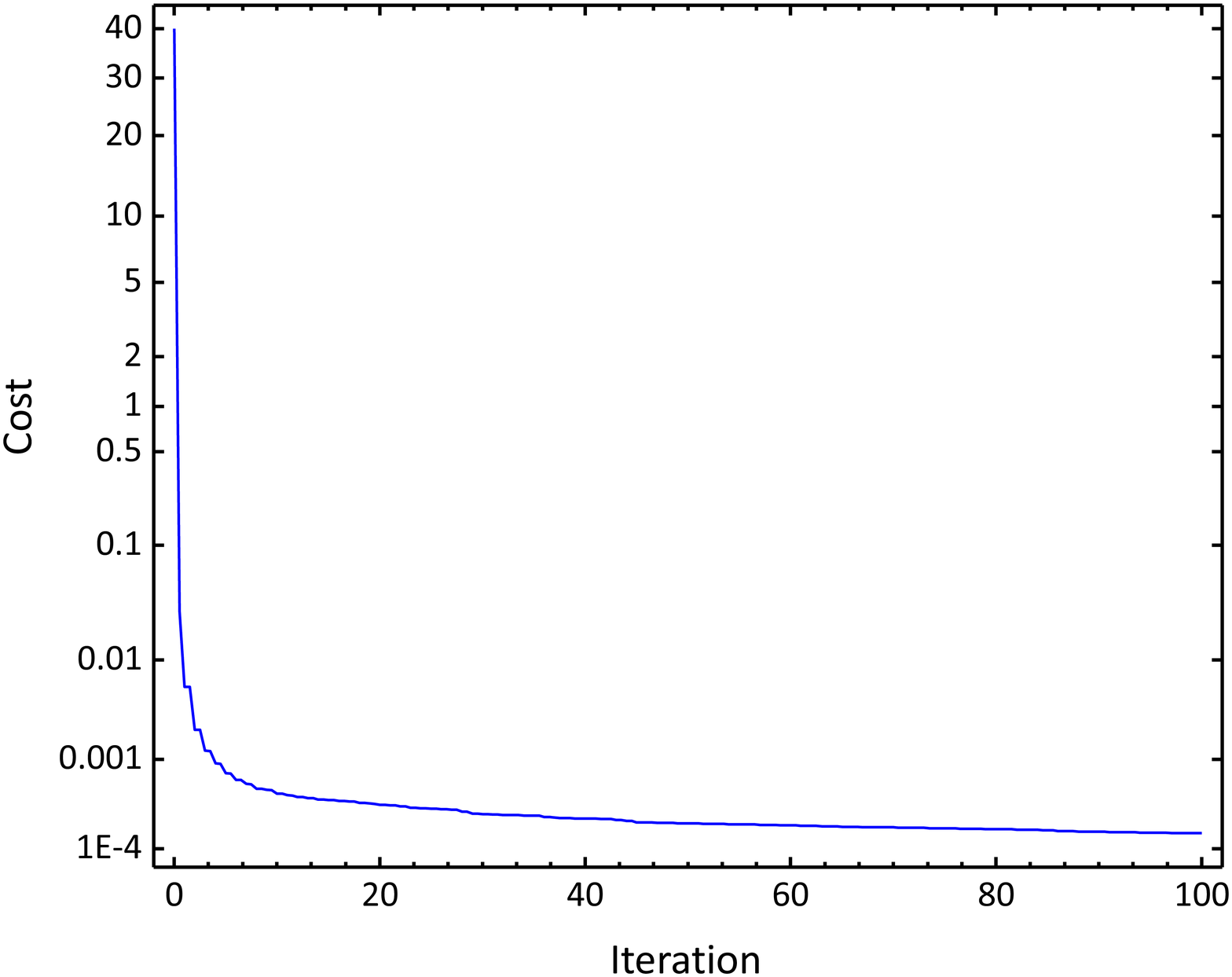}
\caption{
The variation of the cost as the iteration increases. The cost is decreased to near 0 after 100 iterations finished. It means that optimized PMF is very close to Gaussian shape.
} \label{fig:A1}
\end{figure*}

The optimization algorithm is based on MATLAB Global Optimization Toolbox, which runs in a computer with an Intel(R) Core(TM) i9-10900K @3.70GHz CPU, and 64 GB of RAM.
As an example, we choose the following optimization condition: we cut JSA into $N_0=600$ pieces, and the number of domains is $N=1935$. The PSO($\omega_p$, $\sigma_p$) is executed for 100 iterations, and the PSO($\textbf{A}$) is executed for 200 iterations. Then this process is repeated 200 times,  i.e., the total iterations are $(100+200)\times200=60000$ times. PSO is executed for 60000 iterations. The result can be obtained in about 8 hours.

As shown in Fig. \ref{fig:A1}, the cost function falls sharply in the first several iterations and the cost will gradually converge to 0 during the optimization process. This demonstrates the feasibility of our algorithm to the crystal structure design in the MIR range, which involves the optimization of many parameters.

\begin{figure*}[htb]
\centering\includegraphics[width= 0.4\textwidth]{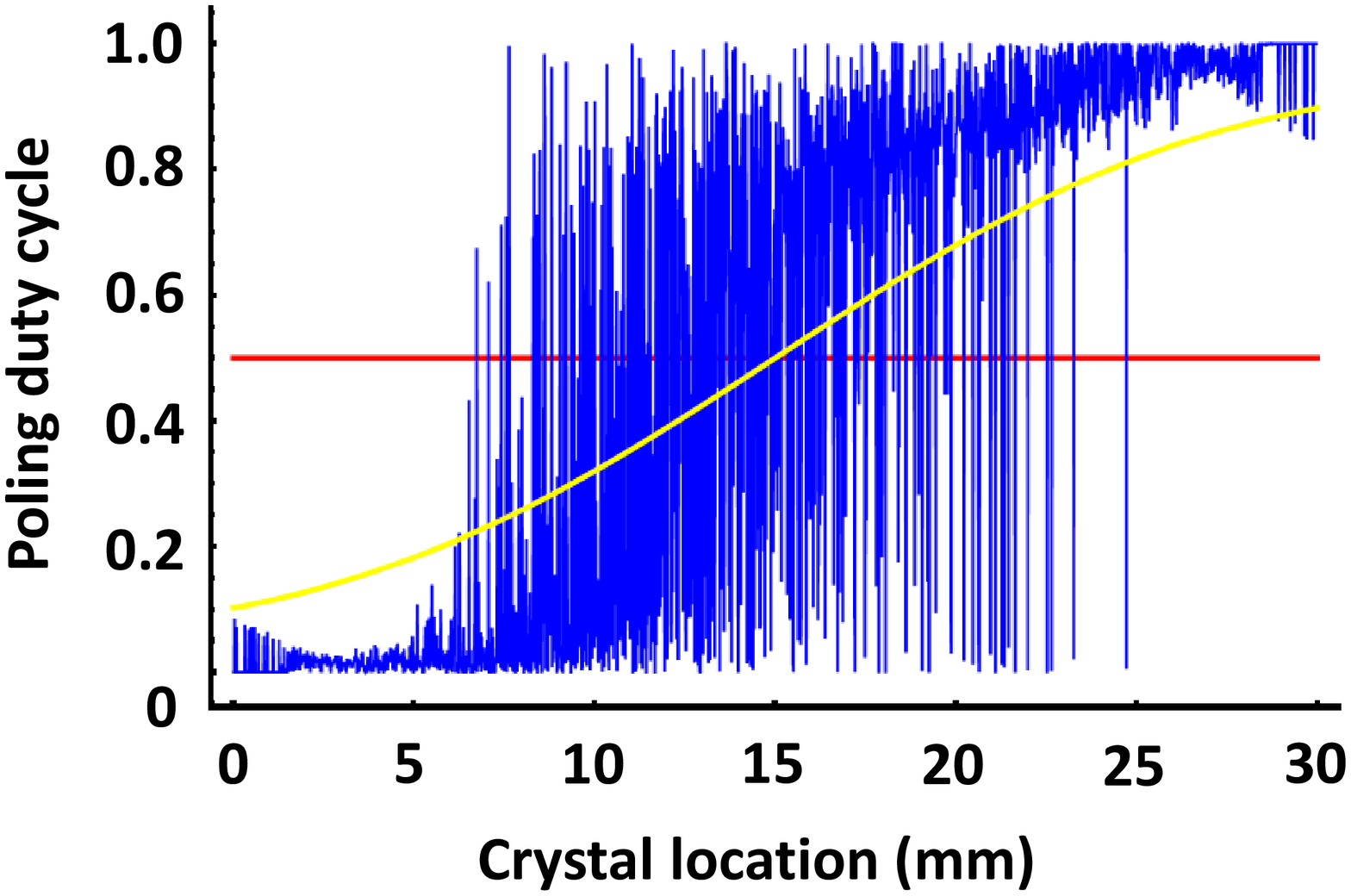}
\caption{
The poling duty cycle \textbf{A} at different positions of the LN crystal. The red line indicates the standard periodically poling, i.e., the duty cycle is always maintained at 50\%. The resultant purity is around 0.82. The yellow line represents the customized poling, i.e., the duty cycle has a Gaussian error function shape. The resultant purity is around 0.864. This is the initial parameters of $\textbf{A}$ in the PSO optimization. The blue line is the optimized poling,  i.e., the duty cycle after PSO optimization.
} \label{fig:A2}
\end{figure*}

After the optimization process is finished, we can get the poling duty cycle \textbf{A} at different positions of the OPLN crystal, as plotted in Fig. \ref{fig:A2}. Meanwhile, we compare the periodically
poling structure of standard PPLN and customized poling LN crystal. We note that the poling structure customized poling LN crystal given by Ref. \cite{Dixon2013} is the initial condition of the optimization process of our work.

For each points in Fig. \ref{fig:3}, the parameters of pump bandwidth, crystal length, purity and the corresponding poling period is presented in Tab. \ref{table:A1}. Unlike other wavelength, we utilize 50-mm-long crystal for the simulation condition of 3.8 $\mu$m and 4.0 $\mu$m in order to get better performance.

\begin{table}[tb]
\centering
\begin{tabular}{cccccccc}
\hline \hline
Wavelength ($\mu$m)                      &3.0           &3.2           &3.4         &3.5        &3.6       &3.8         &4.0\\
\hline
Pump bandwidth $\Delta\lambda$ (nm)      &1.9           &1.9           &2.1         &2.3        &2.5       &2.1         &9.2\\
FWHM ($\approx$1.67$\Delta\lambda$) (nm)          &3.2           &3.2           &3.5         &3.8        &4.2       &3.5         &3.85\\
Crystal length (mm)                      &30            &30            &30          &30         &30        &50          &50\\
Poling period  $\Lambda$ ($\mu$m)        &15.2825       &15.448        &15.504      &15.497     &15.472    &15.3665     &15.207\\
Purity                                   &0.996327      &0.996920      &0.997597    &0.997571   &0.998010  &0.997372    &0.996463\\
\hline
\hline
\end{tabular}
\caption{\label{table:A1} Summary of parameters of OPLN for wavelength-degenerated SPDC. The Pump bandwidth $\Delta\lambda$ (nm), FWHM (1.67$\Delta\lambda$), crystal length (mm), poling period  $\Lambda$ ($\mu$m) and purity are listed.}
\end{table}

\begin{figure*}[!htb]
\centering\includegraphics[width= 0.6\textwidth]{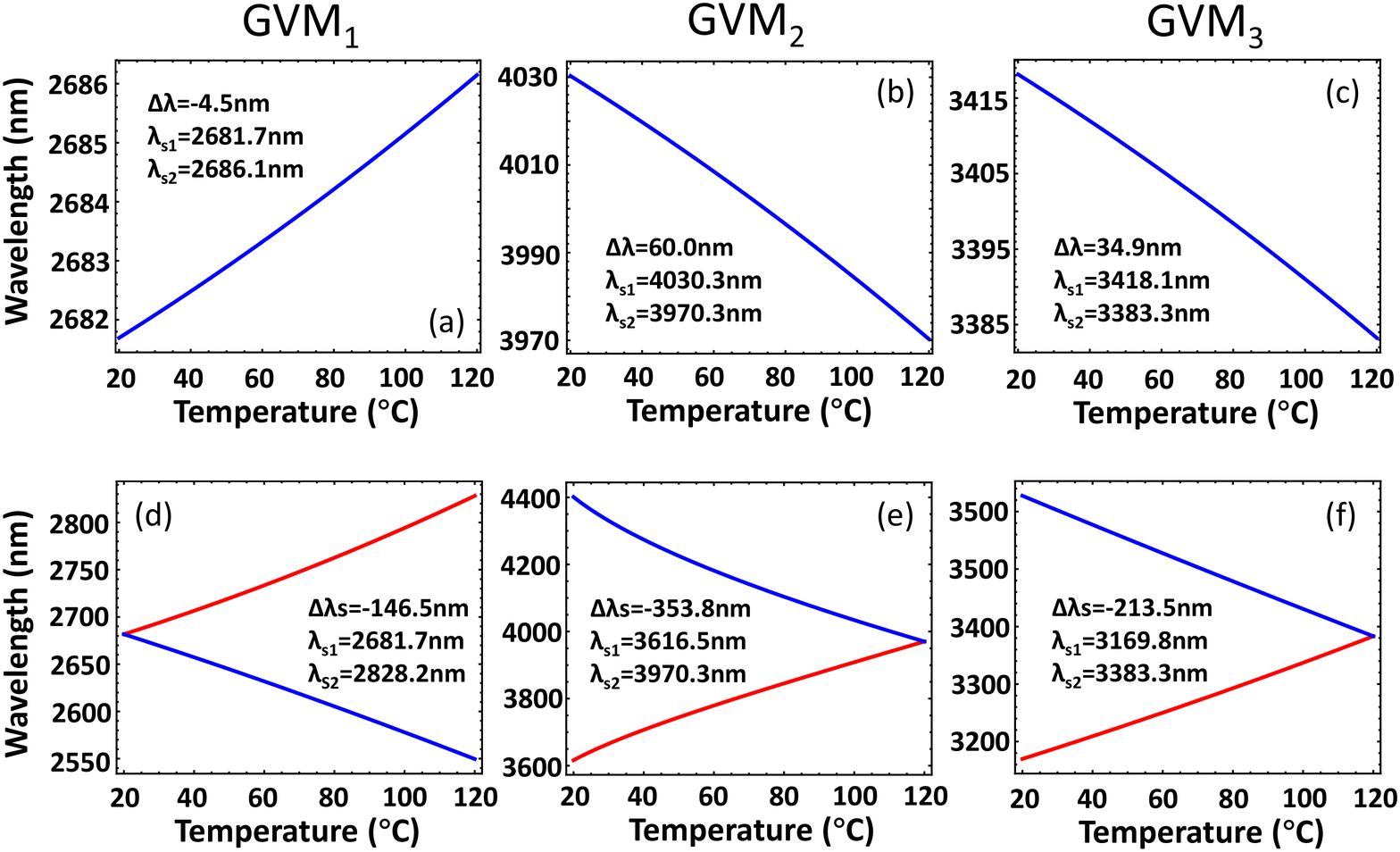}
\caption{
(a)-(c) The three kinds of GVM wavelength with temperature variation from 20 $^\circ$C to 120 $^\circ$C.
(d)-(f) The phase-matched wavelength with temperature dependent for signal (red) and idler (blue).
$\lambda_{1(2)} $ is the wavelength of the signal at 20 (120)$^\circ$C.
 } \label{fig:A3}
\end{figure*}

\subsection*{C. Thermal properties of LN at MIR wavelengths}
In this section, we consider the GVM wavelengths and the phase-matching wavelength at different temperatures.
According to the Sellmeier equations and thermal$-$optical equations \cite{Jundt1990}, we calculate the thermal characteristic of OPLN crystal.
As shown in Fig. \ref{fig:A3} (a)-(c),  the GVM wavelength with the temperature change between $-$4.5 and 60 nm in the range from 20$^\circ$C to 120$^\circ$C.
Fig. \ref{fig:A3} (d)-(f) shows the phase-matching wavelengths under three GVM conditions.
Noted that all the signal is increasing while all the idler is decreasing.
The signal wavelength decreases by 146.5 nm, 353.8 nm, and 213.5 nm, respectively, for three GVM conditions when the temperature increased from 20$^\circ$C to 120$^\circ$C.

\subsection*{D. JSAs are shown at wider wavelength range}
\begin{figure*}[!htb]
\centering\includegraphics[width= 0.6\textwidth]{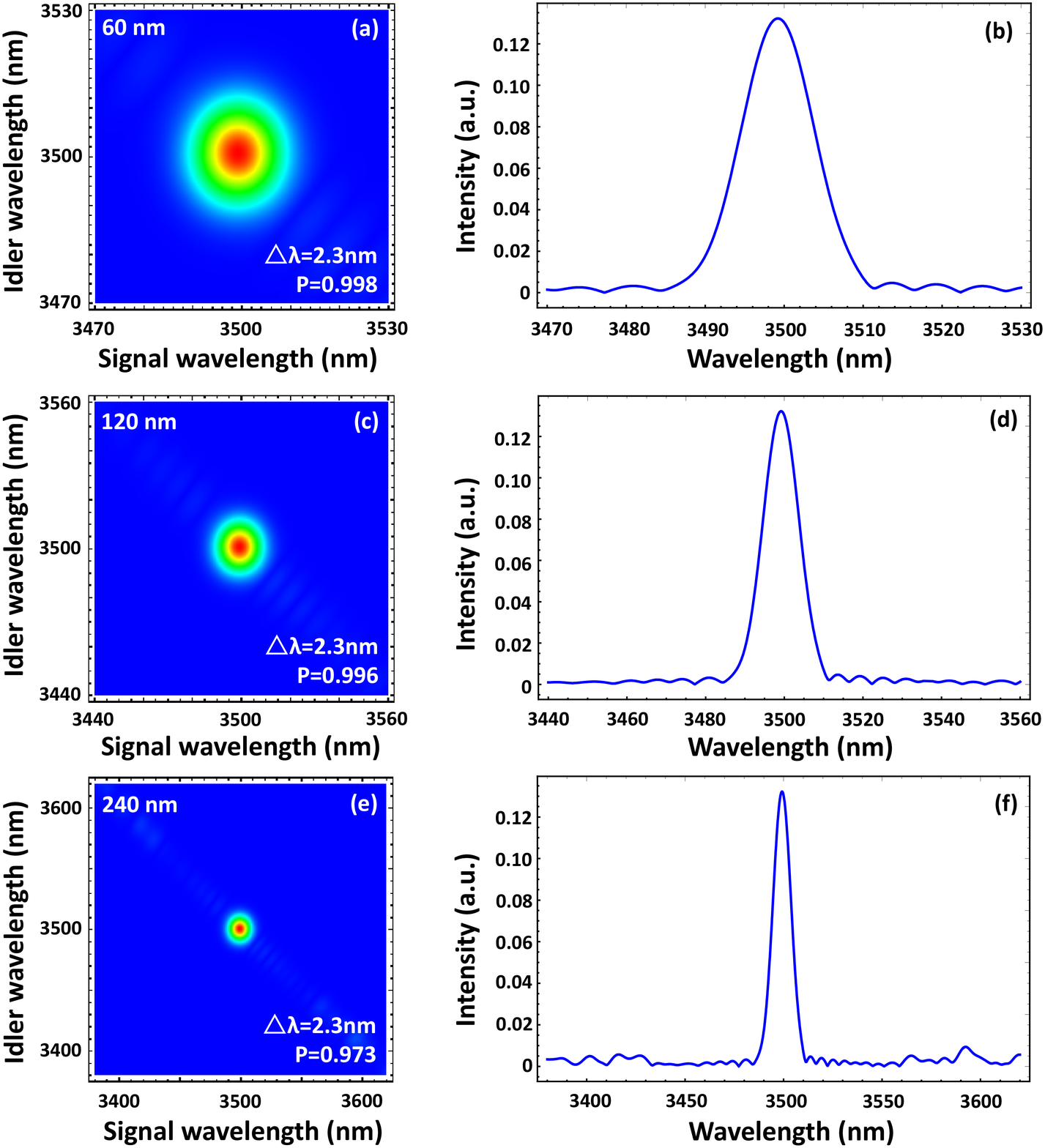}
\caption{(left) JSA at (a) 60 nm, (c) 120 nm, and (e) 240 nm wavelengths range. The parameters of pump bandwidth $\Delta\lambda$, and the purity $P$ are listed in each figure. (right) Projection of JSA on anti-diagonal of (b) 60 nm, (d) 120 nm, and (f) 240 nm  wavelengths range.
} \label{fig:A4}
\end{figure*}

\begin{figure*}[!htb]
\centering\includegraphics[width= 0.5\textwidth]{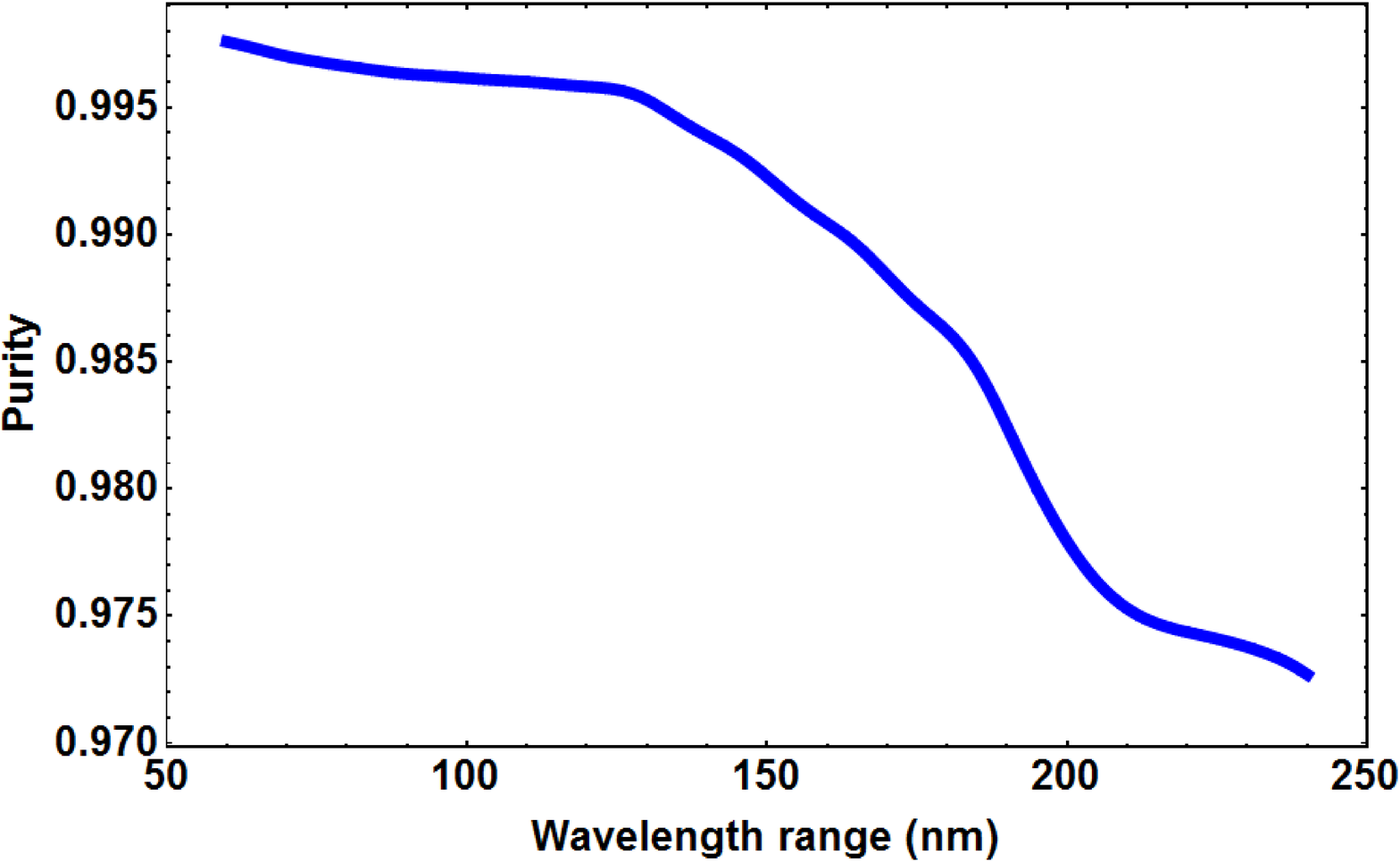}
\caption{Purity for the JSA versus wavelength range. The conditions of 60 nm, 120 nm, and 240 nm correspond to the JSA at Fig. \ref{fig:A4}(a)(c)(e).
} \label{fig:A5}
\end{figure*}

In the main text, all the purities are calculated by considering a wavelength range of 60 nm.
This is equivalent to adding a 60 nm wide square-shape BPF on the source.
Here, we consider the JSA performance at a wider wavelength range.
have depicted PMF projection and JSA in
Figure \ref{fig:A4} shows the JSA and its marginal distribution at the wavelength range of  60 nm, 120 nm, and  240 nm wavelengths range.
It can be observed that when the range is enlarged, more
side-lobes begin to appear, which leads the purity falls down from 0.99 to 0.97.
Figure \ref{fig:A5} shows the calculated purity as a function of wavelength range decreased continuously from 60 nm to 240 nm.
The purity does not decrease rapidly.
This means using a bandpass filter with large width of 240 nm, the purity can still be higher than 0.97.

Besides, we also summarized the wavelength-range-dependent purity variation quantitatively. It will show how the side-lobes in Fig. \ref{fig:A4} influence the purity of JSA specifically.

\newpage
\subsection*{E. Derivation of Equation \ref{eq:5}}
In  this section, we provide the details in the derivation of Eq. (5).
We consider the PMF ($\phi\left(\lambda_{s}, \lambda_{i}\right)_{l}$) of  one domain ( the $l^{th}$ domain), with the position from $z_{l}$ to $z_{l+1}$.
The nonlinear polarization parameter g(z) is defined as
\begin{equation}
g(z) = ( - 1)^l ,\begin{array}{*{20}c}
   {}  \\
\end{array}z_l  < z < z_{l + 1}.
\end{equation}
The PMF can be calculated as
\begin{equation}\label{PMF1}
\begin{array}{lll}
\phi\left(\lambda_{s}, \lambda_{i}\right)_{l}
&=&\frac{1}{L} \int_{z_{l}}^{z_{l+1}} d z g(z) \exp (-i \Delta k z)\\ \\
&=&\left.\frac{i}{\Delta k L}(-1)^{l} e^{-i \Delta k z}\right|_{z_{l}} ^{z_{l+1}}\\ \\
&=&\frac{i}{\Delta kL}(-1)^{l} (e^{-i \Delta k z_{l+1}}-e^{-i \Delta k z_l})\\ \\
&=&\frac{i}{\Delta kL}(-1)^{l}e^{-i \Delta k (z_{l+1}+z_{l})/2} [e^{-i \Delta k (z_{l+1}-z_{l})/2}-e^{i \Delta k (z_{l+1}-z_{l})/2}]\\ \\
&=&\frac{i}{\Delta kL}(-1)^{l}e^{-i \Delta k (z_{l+1}+z_{l})/2} [-2i \sin (\Delta k (z_{l+1}-z_{l})/2)]\\ \\
&=&(-1)^{l}\frac{2}{\Delta kL}e^{-i \Delta k (z_{l+1}+z_{l})/2} \sin (\Delta k (z_{l+1}-z_{l})/2).\\
\end{array}
\end{equation}
By summarizing different $l$,
we obtain

\begin{equation}\label{PMF3}
\begin{aligned}
&\phi\left(\lambda_{s}, \lambda_{i}\right)=
\sum_{l=0}^{N-1}(-1)^{l} \phi\left(\lambda_{s}, \lambda_{i}\right)_l
&=\frac{2}{\Delta k L} \sum_{l=0}^{N-1}\left\{\sin \left[\Delta k\left(z_{l+1}-z_{l}\right) / 2\right] e^{-i \Delta k\left(z_{l+1}+z_{l}\right) / 2}\right\}.
\end{aligned}
\end{equation}

\end{document}